\documentclass[traditabstract]{aa}
\usepackage{graphicx}
\usepackage{txfonts}
\usepackage{natbib}
\bibpunct{(}{)}{;}{a}{}{,}

\newcommand{\Msun}{\ensuremath{\,{\rm M}_\odot}}           
\newcommand{\Rsun}{\ensuremath{\,{\rm M}_\odot}}           
\newcommand{\Teff}{\ensuremath{T_{\rm eff}}}               
\newcommand{\apx}{$^{\prime\prime}$\,px$^{-1}$}            
\newcommand{\Porb}{\ensuremath{P_{\rm orb}}}               
\newcommand{\mc}[1]{\multicolumn{2}{c}{#1}}
\newcommand{\ml}[1]{\multicolumn{2}{l}{#1}}
\newcommand{\reff}[1]{{#1}}                                

\title{Orbital periods of cataclysmic variables identified by the SDSS.}

\subtitle{VII. Four new eclipsing systems}

\author{John Southworth \inst{1,2} \and C.\ M.\ Copperwheat \inst{2}
        \and B.\ T.\ G\"ansicke \inst{2} \and S.\ Pyrzas \inst{2,3} }

\institute{Astrophysics Group, Keele University, Newcastle-under-Lyme, ST5 5BG, UK \ \ \ \ \ \email{jkt@astro.keele.ac.uk}
           \and Department of Physics, University of Warwick, Coventry, CV4 7AL, UK
           \and Isaac Newton Group of Telescopes, Apartado de Correos 321, E-38700, Santa Cruz de La Palma, Spain
          }

\date{Received ????; accepted ????}       

\abstract{We present photometry of nine cataclysmic variable stars identified by the Sloan Digital Sky Survey, aimed at measuring the orbital periods of these systems. Four of these objects show deep eclipses, from which we measure their orbital periods. The light curves of three of the eclipsing systems are also analysed using the {\sc lcurve} code, and their mass ratios and orbital inclinations determined. SDSS J075059.97$+$141150.1 has an orbital period of $134.1564 \pm 0.0008$ min, making it a useful object with which to investigate the evolutionary processes of cataclysmic variables. SDSS J092444.48$+$080150.9 has a period of $131.2432 \pm 0.0014$ min and is probably magnetic. The white dwarf ingress and egress phases are very deep and short, and there is no clear evidence that this object has an accretion disc. SDSS J115207.00$+$404947.8 and SDSS J152419.33$+$220920.1 are nearly identical twins, with periods of $97.5 \pm 0.4$ and $93.6 \pm 0.5$ min and mass ratios of $0.14 \pm 0.03$ and $0.17 \pm 0.03$, respectively. Their eclipses have well-defined white dwarf and bright spot ingress and egress features, making them excellent candidates for detailed study. All four of the orbital periods presented here are shorter than the 2--3\,hour period gap observed in the known population of cataclysmic variables.}


\begin{document} 

\maketitle 

\section{Introduction}                                                                       \label{sec:intro}

Cataclysmic variables (CVs) are interacting binary stars containing a white dwarf primary component in a close orbit with a low-mass secondary star which fills its Roche lobe. In most of these systems the secondary component is hydrogen-rich and transfers material to the white dwarf via an accretion disc. Comprehensive reviews of the properties of CVs have been given by \citet{Warner95book} and \citet{Hellier01book}. CVs are thought to evolve from longer to shorter orbital periods through the loss of orbital angular momentum, before reaching a minimum period caused by the changes in the structure of the mass donor and evolving back to longer periods. However, theoretical population synthesis models of this scenario \citep{Dekool92aa, DekoolRitter93aa, Kolb93aa, KolbDekool93aa, Politano96apj, Politano04apj, KolbBaraffe99mn} are unable to reproduce the orbital period distribution of the observed population of CVs \citep{Downes+01pasp,RitterKolb03aa}.

This disagreement may mainly be due to observational selection effects favouring the discovery and analysis of longer-period CVs, which are intrinsically much brighter than their shorter-period colleagues. We are therefore working on the characterisation of the population of CVs discovered by the Sloan Digital Sky Survey (SDSS\footnote{\tt http://www.sdss.org/}; see \citealt{Szkody+07aj} and references therein), which were identified spectroscopically so should be much less affected by these selection biases. Further information and previous results can be found in \citet{Gansicke+06mn}, \citet{Me+06mn,Me+07mn,Me+07mn2,Me+08mn,Me++08mn} and \citet{Dillon+08mn}. A major recent result of our project is the identification of the long-predicted pile-up of objects at the minimum period (the `period spike'), which is discussed in detail by \citet{Gansicke+09mn}.

\begin{figure} \includegraphics[width=0.48\textwidth,angle=0]{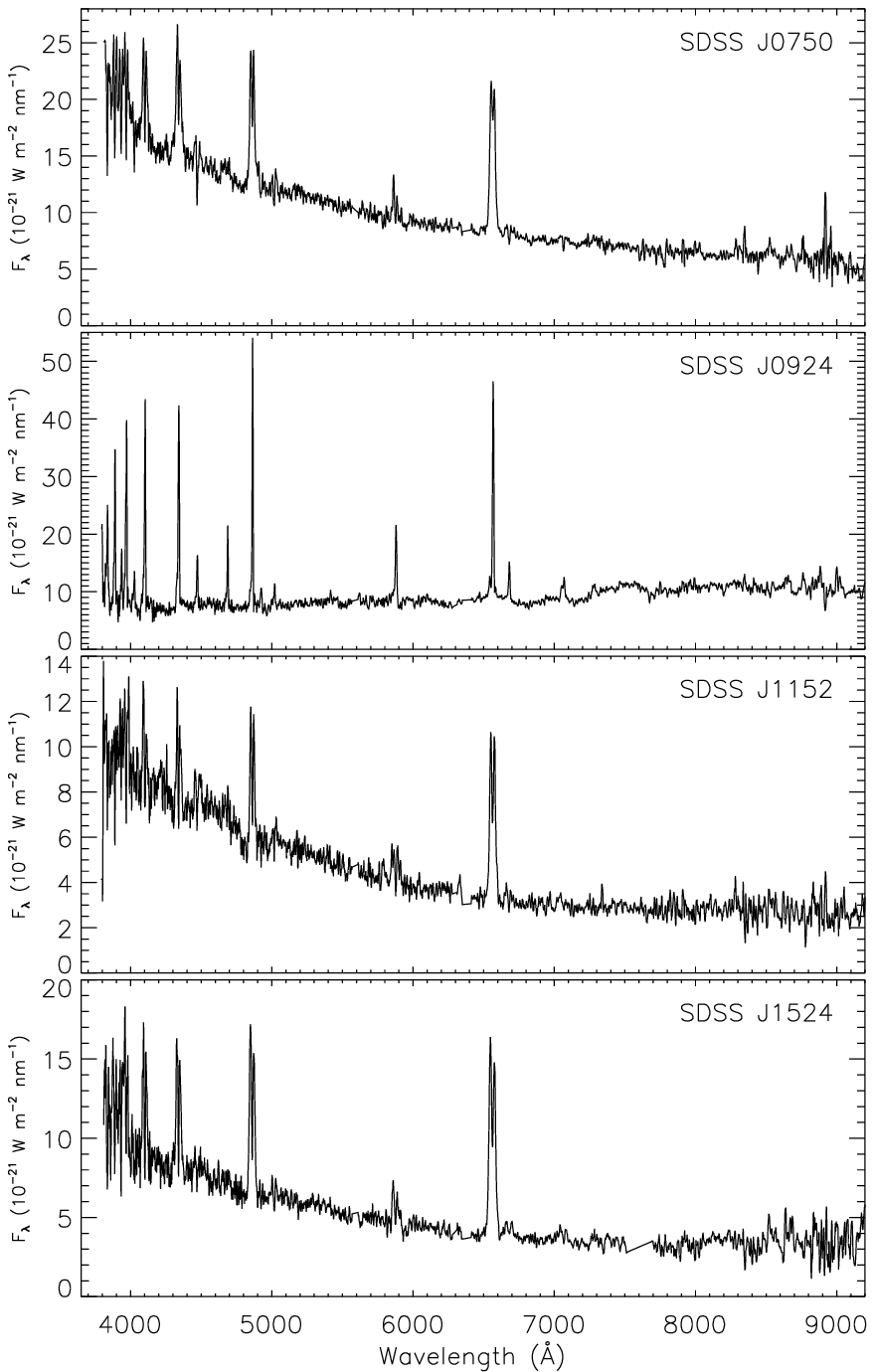} \\
\caption{\label{fig:sdssspec-e} SDSS spectra of the CVs for which we present orbital period
measurements. For this plot the flux levels have been smoothed with 10-pixel Savitsky-Golay
filters. The units of the abscissae are $10^{-21}$\,W\,m$^{-2}$\,nm$^{-1}$, which corresponds
to $10^{-17}$\,erg\,s$^{-1}$\,cm$^{-2}$\,\AA$^{-1}$.} \end{figure}

In this work we present and analyse time-resolved photometry of four CVs which we have discovered to show eclipses. We abbreviate the names of the targets to SDSS\,J$nnnn$, where $nnnn$ are the first four digits of right ascension. The full and abbreviated names, references and $ugriz$ apparent magnitudes are given in Table\,\ref{tab:iddata}. In Fig.\,\ref{fig:sdssspec-e} we plot their SDSS spectra for reference. In Sect.\,\ref{sec:no} we also present light curves of five SDSS CVs for which we did not find periodic brightness variations.

\begin{table*} \begin{center}
\caption{\label{tab:iddata} Apparent magnitudes of our targets in the SDSS $ugriz$
passbands. $g_{\rm spec}$ are apparent magnitudes we have calculated by convolving
the SDSS flux-calibrated spectra with the $g$ passband function. They are obtained
at a different epoch to the SDSS $ugriz$ magnitudes, but are affected by `slit losses'
and any errors in astrometry or positioning of the spectroscopic fibre entrance.}
\begin{tabular}{lllccccccc} \hline
SDSS name                  & Short name  & Reference           & $u$ & $g$ & $r$ & $i$ & $z$ & $g_{\rm spec}$  \\
\hline
SDSS J031051.66$-$075500.2 & SDSS\,J0310 & \citet{Szkody+03aj} & 15.76 & 15.49 & 15.74 & 15.89 & 16.10 & 22.03 \\
SDSS J074355.55$+$183834.8 & SDSS\,J0743 & \citet{Szkody+06aj} & 20.11 & 20.07 & 19.29 & 18.78 & 18.61 & 21.66 \\
SDSS J074640.61$+$173412.7 & SDSS\,J0746 & \citet{Szkody+06aj} & 18.21 & 18.16 & 18.37 & 18.52 & 18.55 & 21.20 \\
SDSS J075059.97$+$141150.1 & SDSS\,J0750 & \citet{Szkody+07aj} & 19.20 & 19.09 & 18.98 & 18.79 & 18.58 & 18.83 \\
SDSS J075808.81$+$104345.5 & SDSS\,J0758 & \citet{Szkody+09aj} & 17.06 & 16.92 & 17.07 & 17.17 & 17.25 & 17.06 \\
SDSS J090628.24$+$052656.9 & SDSS\,J0906 & \citet{Szkody+05aj} & 18.82 & 18.76 & 18.46 & 18.10 & 17.82 & 19.32 \\
SDSS J092444.48$+$080150.9 & SDSS\,J0924 & \citet{Szkody+05aj} & 19.49 & 19.25 & 19.26 & 18.70 & 17.88 & 19.36 \\
SDSS J115207.00$+$404947.8 & SDSS\,J1152 & \citet{Szkody+07aj} & 19.22 & 19.26 & 19.15 & 19.13 & 19.00 & 19.56 \\
SDSS J152419.33$+$220920.1 & SDSS\,J1524 & \citet{Szkody+09aj} & 19.04 & 19.04 & 18.90 & 18.80 & 18.50 & 19.52 \\
\hline \end{tabular} \end{center} \end{table*}

\begin{table*} \begin{center}
\caption{\label{tab:obslog} Log of the observations presented in this work.
The mean magnitudes are formed from observations outside eclipses only.}
\begin{tabular}{lcccccccc} \hline
Target    & Date & Start time & End time & Telescope and & Optical &  Number of   & Exposure & Mean      \\
          & (UT) &  (UT)      &  (UT)    &  instrument   & element & observations & time (s) & magnitude \\
\hline
SDSS\,J0310 & 2009 01 25 & 00:56 & 03:40 & NTT\,/\,EFOSC2 & $B_{\rm Tyson}$ filter & 105 &     60 & 21.3 \\
[2pt]
SDSS\,J0743 & 2009 01 23 & 01:36 & 03:37 & NTT\,/\,EFOSC2 & $B_{\rm Tyson}$ filter &  80 & 40--60 & 22.3 \\
SDSS\,J0743 & 2009 01 23 & 05:39 & 06:43 & NTT\,/\,EFOSC2 & $B_{\rm Tyson}$ filter &  42 &     60 & 22.2 \\
[2pt]
SDSS\,J0746 & 2009 01 24 & 01:01 & 03:44 & NTT\,/\,EFOSC2 & $B_{\rm Tyson}$ filter & 123 & 30--60 & 20.2 \\
SDSS\,J0746 & 2009 01 24 & 06:09 & 07:12 & NTT\,/\,EFOSC2 & $B_{\rm Tyson}$ filter &  59 &     30 & 20.4 \\
[2pt]
SDSS\,J0750 & 2009 01 22 & 00:53 & 05:27 & NTT\,/\,EFOSC2 & $B_{\rm Tyson}$ filter & 255 &     30 & 19.2 \\
SDSS\,J0750 & 2009 01 23 & 01:07 & 01:30 & NTT\,/\,EFOSC2 & $B_{\rm Tyson}$ filter &  22 &     30 & 19.3 \\
SDSS\,J0750 & 2009 01 26 & 02:50 & 03:33 & NTT\,/\,EFOSC2 & $B_{\rm Tyson}$ filter &  41 &     30 & 19.4 \\
SDSS\,J0750 & 2009 01 27 & 00:57 & 01:37 & NTT\,/\,EFOSC2 & $B_{\rm Tyson}$ filter &  34 &     40 & 19.5 \\
SDSS\,J0750 & 2009 02 16 & 22:04 & 22:31 & WHT\,/\,Aux    & $V$ filter             & 164 &      5 & 18.9 \\
[2pt]
SDSS\,J0758 & 2009 01 26 & 00:55 & 02:46 & NTT\,/\,EFOSC2 & $B_{\rm Tyson}$ filter & 150 &     10 & 17.3 \\
SDSS\,J0758 & 2009 01 26 & 04:37 & 04:57 & NTT\,/\,EFOSC2 & $B_{\rm Tyson}$ filter &  28 &     10 & 17.3 \\
[2pt]
SDSS\,J0906 & 2008 01 18 & 04:07 & 06:12 & INT\,/\,WFC    & $g$ filter             &  69 &     70 & 18.9 \\
SDSS\,J0906 & 2008 01 20 & 03:15 & 06:01 & INT\,/\,WFC    & $g$ filter             & 112 &     60 & 18.4 \\
SDSS\,J0906 & 2009 01 24 & 03:49 & 06:04 & NTT\,/\,EFOSC2 & $B_{\rm Tyson}$ filter & 125 &     30 & 19.4 \\
SDSS\,J0906 & 2009 01 24 & 07:16 & 09:17 & NTT\,/\,EFOSC2 & $B_{\rm Tyson}$ filter & 113 &     30 & 19.4 \\
SDSS\,J0906 & 2009 01 27 & 01:43 & 03:28 & NTT\,/\,EFOSC2 & $R$ filter             &  85 &     40 & 19.4 \\
[2pt]
SDSS\,J0924 & 2009 01 25 & 03:46 & 07:29 & NTT\,/\,EFOSC2 & $B_{\rm Tyson}$ filter & 207 &     30 & 21.1 \\
SDSS\,J0924 & 2009 01 26 & 05:00 & 05:21 & NTT\,/\,EFOSC2 & $B_{\rm Tyson}$ filter &  20 &     30 & 21.4 \\
SDSS\,J0924 & 2009 01 27 & 05:04 & 05:34 & NTT\,/\,EFOSC2 & $B_{\rm Tyson}$ filter &  29 &     30 & 21.3 \\
SDSS\,J0924 & 2009 02 16 & 23:51 & 00:10 & WHT\,/\,Aux    & $V$ filter             &  75 &     10 & 18.9 \\
[2pt]
SDSS\,J1152 & 2009 02 17 & 00:16 & 02:36 & WHT\,/\,Aux    & $V$ filter             & 564 &     10 & 19.6 \\
[2pt]
SDSS\,J1524 & 2009 05 06 & 02:58 & 05:43 & WHT\,/\,Aux    & $V$ filter             & 756 &     10 & 19.1 \\
\hline \end{tabular} \end{center} \end{table*}

\section{Observations and data reduction}                                                      \label{sec:obs}

Most of the observations presented in this work were obtained in 2009 January using the New Technology Telescope (NTT) at ESO La Silla, equipped with the EFOSC2 focal-reducing instrument%
\footnote{\tt http://www.eso.org/sci/facilities/lasilla/ instruments/efosc/index.html}
\citep{Buzzoni+84msngr}. We used EFOSC2 in imaging mode and with a Loral 2048$\times$2048\,px$^2$ CCD binned 2$\times$2, giving a field of view of 4.4$\times$4.4\,arcmin$^2$ at a plate scale of 0.26\apx. All images were obtained with a $B_{\rm Tyson}$ filter (ESO filter \#724), which has a central wavelength of 4445\,\AA\ and a FWHM of 1838\,\AA.

Additional observations were obtained in 2009 February (classical observing) and May (service mode) at the William Herschel Telescope (WHT) at La Palma. For these we used the Auxiliary Port focal station, which was equipped at the time with the EEV 4200$\times$2048\,px$^2$ CCD destined for the ACAM instrument%
\footnote{\tt http://www.ing.iac.es/Astronomy/instruments/acam/}.
The CCD was binned 4$\times$4 and windowed down to the central 2048$\times$2100 pixels, which reduced the dead time between exposures to only 5\,s whilst still properly sampling the seeing conditions (plate scale 0.27\apx) and available field of view (a circle of 1.8\,arcmin in diameter). We used the Johnson $V$ filter for all WHT observations.

The data were reduced using standard methods, with a pipeline written in {\sc idl}%
\footnote{The acronym {\sc idl} stands for Interactive Data Language and is a trademark of ITT Visual
Information Solutions. For further details see {\tt http://www.ittvis.com/ProductServices/IDL.aspx}.}
 \citep{Me+09mn2,Me+09apj,Me+09mn}. Aperture photometry was performed using the {\sc astrolib}/{\sc aper} procedure%
\footnote{The {\sc astrolib} subroutine library is distributed by NASA. For further details see {\tt http://idlastro.gsfc.nasa.gov/}.},
which originates from {\sc daophot} \citep{Stetson87pasp}. For the NTT we were able to set the photometry apertures by eye and fix their positions, whereas for the WHT they were set by eye and adjusted according to the offset between each image and the reference image found by cross-correlation.

Light curves of SDSS\,J0906 were obtained on the nights of 2008/01/17 and 2008/01/19 using the Isaac Newton Telescope (INT) and Wide Field Camera (WFC) equipped with a $g$ filter. These data were reduced using the pipeline described by \citet{gansicke+04aa}, which performs aperture photometry via the {\sc SExtractor} package \citep{BertinArnouts96aas}.

The instrumental differential magnitudes have been transformed to $BV$ apparent magnitudes using formulae from \citet{Jordi++06aa}. Given the non-standard nature of the spectral energy distribution of CVs, and the $B_{\rm Tyson}$ filter used at the NTT, the resulting apparent magnitudes are uncertain by at least a few tenths of a magnitude.


\section{Light curve modelling}                                                             \label{sec:lcurve}

\begin{table*} \centering
\caption{\label{tab:lcurve} Parameter determinations from our {\sc lcurve} fits
for SDSS\,J0750, SDSS\,J1152 and SDSS\,J1524. The radius of the WD and the radial
position of the bright spot are given as fractions of the orbital semimajor axis, $a$.}
\begin{tabular}{l l l r@{\,$\pm$\,}l r@{\,$\pm$\,}l r@{\,$\pm$\,}l} \hline
\ml{Parameter}                      &            & \mc{SDSS\,J0750}  & \mc{SDSS\,J1152}  & \mc{SDSS\,J1524} \\
\hline
$q$          & Mass ratio           &            & $0.59$  & $0.17$  & $0.14$  & $0.03$  & $0.17$  &$0.03$  \\
$i$          & Orbital inclination  & ($^\circ$) & $74.6$  & $1.9$   & $83.7$  & $1.8$   & $82.8$  &$0.8$   \\
$T_0$ & Eclipse midpoint&(HJD)&$2454853.62917$&$0.00004$&$2454879.53815$&$0.00001$&$2454957.64961$&$0.00003$\\
$r_{\rm WD}$ & Radius of the WD     & ($a$)      & $0.020$ & $0.004$ & $0.012$ & $0.002$ & $0.034$ &$0.002$ \\
$r_{\rm BS}$ & Bright spot position & ($a$)      & $0.16$  & $0.08$  & $0.31$  & $0.04$  & $0.29$  &$0.02$  \\
\hline \end{tabular} \end{table*}

We have modelled the observed eclipses of SDSS\,J0750, SDSS\,J1152 and SDSS\,J1524 in order to explore the physical properties of these systems. For this analysis we used {\sc lcurve}, a code developed to fit light curves of eclipsing binary stars containing a degenerate object. A complete description of {\sc lcurve} is given in \citet{Copperwheat+09}. The binary is defined by four components: white dwarf, secondary star, accretion disc and bright spot. We obtained initial fits using the Nelder-Mead downhill simplex and Levenberg-Marquardt methods \citep{Press+92book}. We then used a Markov Chain Monte Carlo (MCMC) algorithm for final minimisation and determination of uncertainties \citep[see][]{Copperwheat+09}.

We list the parameter determinations from these fits in Table\,\ref{tab:lcurve}. We include the following quantities: mass ratio ($q$), orbital inclination ($i$), time of white dwarf mid-eclipse ($T_0$), and the white dwarf radius and bright spot position expressed as fractions of the orbital semi-major axis ($r_{\rm WD}$ and $r_{\rm BS}$). We use the bright spot position as an indicator of the accretion disc radius, since we assume that the bright spot lies at the outer edge of the accretion disc.

{\sc lcurve} \reff{implements} a wealth of additional parameters which are needed for modelling high-precision data. The discovery data presented in this work are insufficient to constrain these parameters, so we fixed them at values which we have found to be physically appropriate in detailed modelling of other eclipsing systems \citep{Copperwheat+09, Pyrzas+09mn, Nebot+09aa, Me+09aa}. Our fits are therefore only indicative, and definitive results will require high-quality follow-up light curves. We discuss the results for each system below.

\begin{table} \begin{center}
\caption{\label{tab:eclipses} Times of eclipse for the objects studied
in this work, and the residuals with respect to the calculated linear
ephemerides. The timings below were measured using the simple mirror-image
method (Sect.\,\ref{sec:0750}) in order to maximise their archival value.}
\begin{tabular}{l r r@{\,$\pm$\,}l r} \hline
Object & Cycle & \mc{Time of eclipse (HJD)}  & Residual  \\
\hline
SDSS\,J0750 &   0.0 & 2454853.6293 & 0.0002 & -0.00003 \\
SDSS\,J0750 &   1.0 & 2454853.7225 & 0.0002 &  0.00001 \\
SDSS\,J0750 &  10.0 & 2454854.5612 & 0.0003 &  0.00023 \\
SDSS\,J0750 &  43.0 & 2454857.6353 & 0.0002 & -0.00009 \\
SDSS\,J0750 &  53.0 & 2454858.5670 & 0.0003 & -0.00003 \\
SDSS\,J0750 & 277.0 & 2454879.4358 & 0.0001 &  0.00000 \\[2pt]
SDSS\,J0924 &   0.0 & 2454856.7192 & 0.0002 & -0.00013 \\
SDSS\,J0924 &   1.0 & 2454856.8104 & 0.0003 & -0.00007 \\
SDSS\,J0924 &  11.0 & 2454857.7220 & 0.0002 &  0.00011 \\
SDSS\,J0924 &  22.0 & 2454858.7245 & 0.0002 &  0.00006 \\
SDSS\,J0924 & 250.0 & 2454879.5046 & 0.0002 & -0.00001 \\[2pt]
SDSS\,J1152 &   0.0 & 2454879.5388 & 0.0002 &          \\
SDSS\,J1152 &   1.0 & 2454879.6065 & 0.0002 &          \\[2pt]
SDSS\,J1524 &   0.0 & 2454957.6499 & 0.0001 &          \\
SDSS\,J1524 &   1.0 & 2454957.7149 & 0.0003 &          \\
\hline \end{tabular} \end{center} \end{table}


\section{SDSS J075059.97$+$141150.1}                                                            \label{sec:0750}

\begin{figure} \includegraphics[width=0.48\textwidth,angle=0]{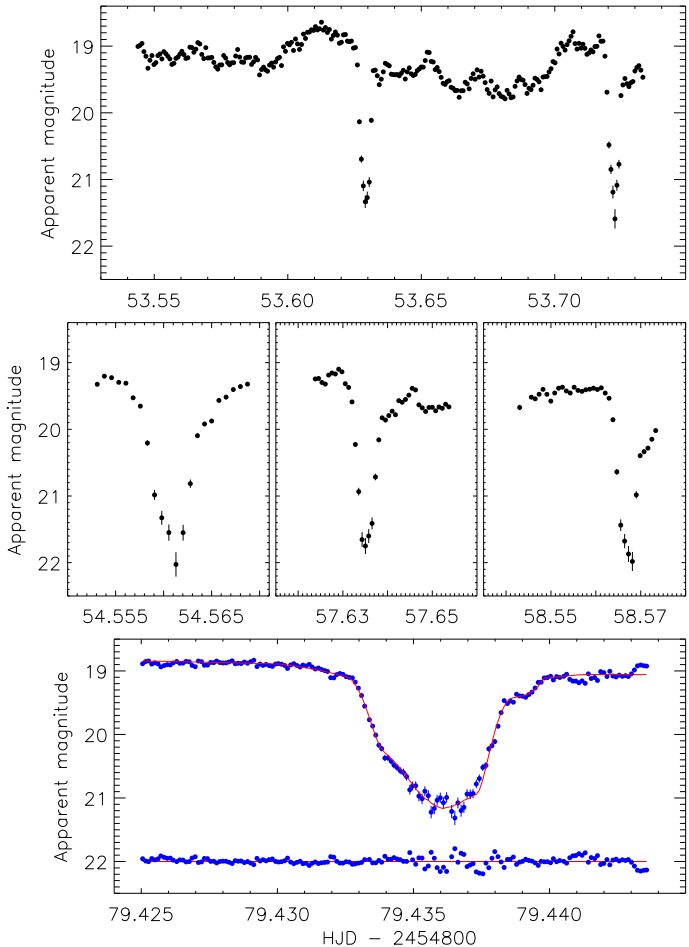} \\
\caption{\label{fig:0750} Light curves of SDSS\,J0750. NTT data ($B_{\rm Tyson}$
filter) are shown on all panels except the bottom one, which contains the WHT data
($V$ filter). The solid line represents our best {\sc lcurve} fit to the WHT data,
and the residuals of the fit are plotted offset from zero for clarity. Note that
the time axis has been stretched to suit each panel.} \end{figure}

SDSS\,J0750 was identified by \citet{Szkody+07aj} as a CV from its SDSS spectrum, which shows a blue continuum with broad and double-peaked hydrogen Balmer and \ion{He}{I} emission lines (Fig.\,\ref{fig:sdssspec-e}). To our knowledge no other information is available in the literature for this object beyond a few catalogued magnitude measurements.

We observed SDSS\,J0750 on the night of 2009/01/20 using NTT/EFOSC2, detecting two eclipses in 4.5\,hr of observing (Fig.\,\ref{fig:0750}). It was in fact undergoing an eclipse whilst we tried to acquire it, although this was not clear at the time. We observed it again on subsequent nights during the same run, and with the WHT/ACAM a month later, in order to extend the time interval on which our orbital period measurement is based.

For each eclipse, a mirror-image of the light curve was shifted until the respective ascending and descending branches were in the best agreement. The time defining the axis of reflection was taken as the midpoint of the eclipses, and uncertainties were estimated based on how far this could be shifted before the agreement was visually poorer. We have fitted a linear ephemeris to these times of minimum light, finding
\[ {\rm Min\,I\ (HJD}) = 2454853.62932 (11) + 0.09316415 (54) \times E \]
where $E$ is the cycle number and the \reff{parenthesised} quantities indicate the uncertainty in the last digit of the preceding number. This corresponds to an orbital period of $134.1564 \pm 0.0008$\,min. The measured times of minimum light and the observed minus calculated values are given in Table\,\ref{tab:eclipses}.

\subsection{Photometric model for SDSS\,J0750}

We have modelled the WHT/ACAM light curve, which has the highest observing cadence, using the {\sc lcurve} code (Sect.\,\ref{sec:lcurve}). The best fit is shown in Fig.\,\ref{fig:0750} and the parameters of the fit are given in Table\,\ref{tab:lcurve}. The relatively featureless light variation makes this quite a difficult object, and our mass ratio measurement is correspondingly uncertain (Table\,\ref{tab:lcurve}).

The photometric determination of parameters in CVs is dependent on measurement of the phase widths of the eclipses of the white dwarf and bright spot. The phase width of the white dwarf eclipse is intrinsically linked to $i$ and $q$, causing these two parameters to be degenerate \citep{Bailey79mn}. This degeneracy can be broken using the observed ingress and egress phases of the bright spot, but in the SDSS\,J0750 light curve these features are rather weakly defined. This causes a large uncertainty in the position of the bright spot and thus the mass ratio. High-speed photometry of several eclipses should be sufficient to provide a clear detection of the bright spot ingress and egress, and thus obtain a precise mass ratio \citep[e.g.][]{Littlefair+06sci,Littlefair+08mn,Copperwheat+09}.


\section{SDSS J092444.48$+$080150.9}

\begin{figure} \includegraphics[width=0.48\textwidth,angle=0]{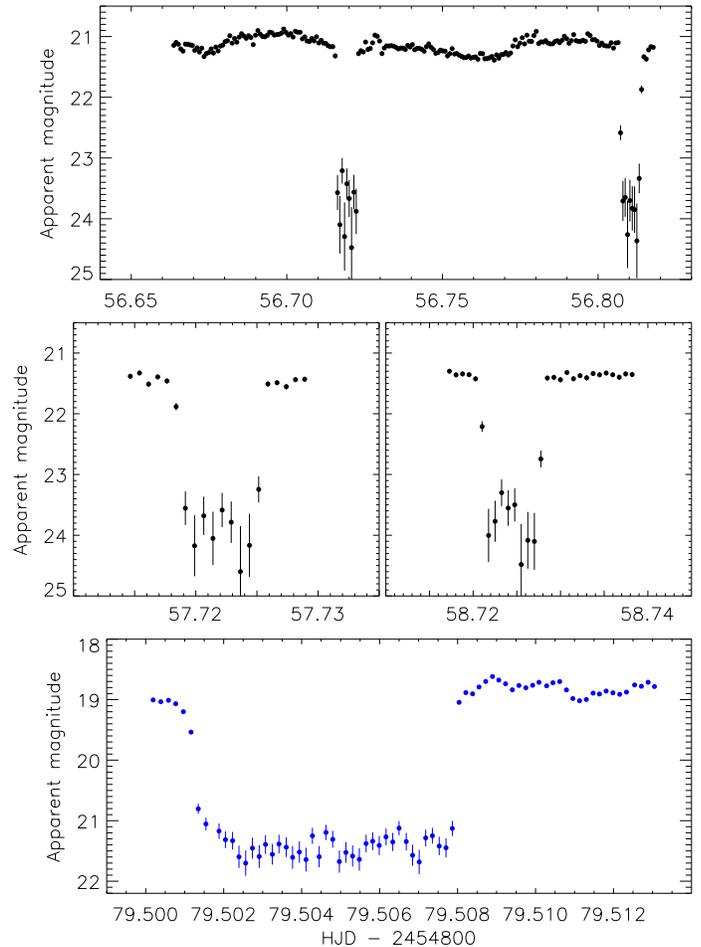} \\
\caption{\label{fig:0924} Light curves of SDSS\,J0924. NTT data ($B_{\rm Tyson}$
filter) are shown on all panels except the bottom one, which contains the WHT data
($V$ filter). The time axis has been stretched to suit each panel.} \end{figure}

SDSS\,J0924 is another new discovery from the SDSS survey, identified by \citet{Szkody+05aj} due to a spectrum which shows strong and narrow Balmer emission and hints of the secondary star towards the red end. The presence of strong \ion{He}{II} is indicative of a magnetic system, but polarimetric observations revealed no polarisation. \citet{Szkody+05aj} presented follow-up observations comprising 11 spectra, which indicated an orbital period in the region of 2.1\,hr, and 3.75\,hr of photometry which showed clear variability but no eclipses.

We observed SDSS\,J0924 on the night of 2009/01/24 using the NTT, obtaining a light curve covering 3.75\,hr. These data cover two eclipses, which are approximately 2.5\,mag deep and 10\,min long (Fig.\,\ref{fig:0924}). We observed one additional eclipse on each of the two following nights, plus one more on the night of 2009/02/16 with the WHT. The eclipse midpoints were obtained as above and a straight line fitted to them to determine an ephemeris:
\[ {\rm Min\,I\ (HJD}) = 2454856.71933 (11) + 0.09114110 (94) \times E \]
which equates to an orbital period of $131.2432 \pm 0.0014$\,min. The measured times of minimum light and the observed minus calculated values are given in Table\,\ref{tab:eclipses}.

The ascending and descending branches of the eclipses of SDSS\,J0924 are extremely quick and are not resolved by our data, even though the WHT observations have a mean cadence of 15.2\,s. The dominant light source in the system is therefore very small, which requires it to be all or part of the white dwarf. The Balmer emission lines indicate that there is mass transfer occuring in this system, but the eclipse morphology indicates that any accretion disc is rather faint. This is clear evidence that the white dwarf is magnetic and its field is disrupting the disc. This system is very well suited to the measurement of the radius of the white dwarf, although this would need high-speed photometry obtained using a large telescope. We did not attempt to model the light curve; our data do not resolve the eclipse ingress and egress shapes so the {\sc lcurve} solutions would be indeterminate.

It is not clear why eclipses were not detected by \citet{Szkody+05aj}, although it must be remembered that the telescope available to these authors was much smaller (1.0\,m versus 3.6\,m). Dr.\ A.\ Henden (private communication) has kindly supplied the data from Szkody et al.\ for a closer inspection. We find that their light curve closely matches the morphology of our own data from the night of 2009/01/24, and that it is possible to align the two eclipses we observed on that night with two short gaps in Szkody et al.'s data. Their non-detection therefore can be attributed to either bad luck with the observing conditions or equipment, or to accidental rejection of data where SDSS\,J0924 was in eclipse and thus below the detection limit of individual images.


\subsection{Spectral energy distribution of SDSS\,J0924}

\begin{figure} \centering
\includegraphics[angle=-90,width=\columnwidth]{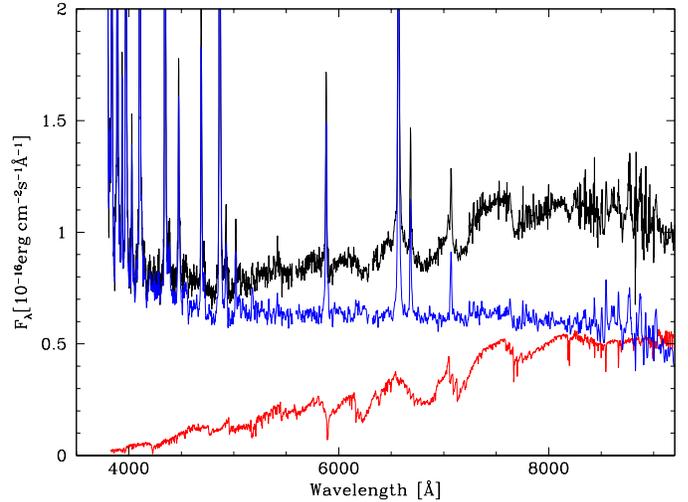}
\caption{\label{fig:0924:sec} Black line: the SDSS spectrum of SDSS\,J0924.
Red line: an M2 template spectrum scaled to match the strengths of the spectral
features of the secondary star. Blue line: the residual spectrum obtained after
subtracting the M-dwarf template from the spectrum of SDSS\,J0924.} \end{figure}

\begin{figure} \centering
\includegraphics[angle=-90,width=\columnwidth]{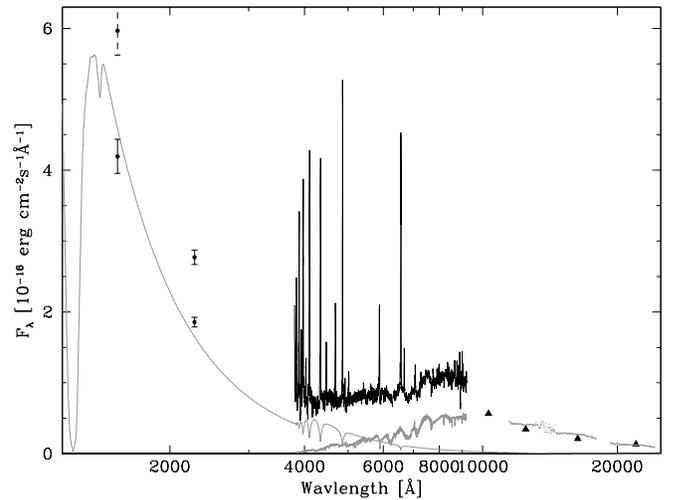}
\caption{\label{fig:0924:sed} Black line: the SDSS spectrum of SDSS\,J0924.
Black dots: ultraviolet fluxes of SDSS\,J0924 from the GALEX Medium Imaging Survey
(lower pair) and dereddened by the full galactic dust column \citep{Schlegel++98apj},
$E_{B-V} = 0.047$ (upper pair). Black triangles: UKIDSS $yJHK$ fluxes. Gray lines:
best-fitting M2 secondary star spectrum from Fig.\,\ref{fig:0924:sec} and a white
dwarf with $\Teff = 17\,000$\,K and $\log g=8.3$, scaled for $d=650$\,pc, as a
plausible match to the observed ultraviolet fluxes.} \end{figure}

The secondary star dominates the optical spectrum of SDSS\,J0924 at longer wavelengths, exhibiting potassium and sodium absorption lines as well as strong TiO bandheads characteristic of mid-to-late M dwarfs. We have determined the star's spectral type using an M dwarf template library \citep{Rebassa+07mn}. These templates were scaled and subtracted from the SDSS spectrum of SDSS\,J0924, and the smoothest residual spectrum was obtained for a spectral type of M2$\pm$0.5 (Fig.\,\ref{fig:0924:sec}). Two SDSS spectra are available for this object, and they give consistent results. From this best-fitting template, we calculated $f_\mathrm{TiO}$, the flux difference between the bands $7450$--$7550$\,\AA\ and $7140$--$7190$\,\AA\ \citep{Beuermann06aa}. For the orbital period of SDSS\,J0924, and an assumed mass ratio of $q=0.25$, we find $R_2 = 0.23$\Rsun\ (only weakly dependent on $q$). Using the polynomial expressions of \citet{Beuermann06aa}, we find  a distance of $d = 650\pm30$\,pc.

SDSS\,J0924 has also been detected in the ultraviolet (UV) Medium Imaging Survey by GALEX \citep{Morrissey+07apjs} and in the infrared (IR) by UKIDSS \citep{Lawrence+07mn}. The UV-optical-IR spectral energy distribution is shown in Fig.\,\ref{fig:0924:sed}. For a distance of 650\,pc, the UV fluxes are consistent with a $15\,000$--$20\,000$\,K white dwarf, depending on its surface gravity (and hence mass). This temperature range is entirely consistent with that observed in a number of \reff{magnetic CVs} \citep{TownsleyGansicke09apj}. Assuming a canonical mass for a CV white dwarf, $M_\mathrm{wd}=0.8$\Msun\ and an effective temperature of $17\,000$\,K, we find a good match to the GALEX fluxes and the blue upturn of the SDSS spectrum (Fig.\,\ref{fig:0924:sed}).

For a system at the lower edge of the period gap, \citet{Beuermann+98aa} and \citet{Knigge06mn} suggest that the companion should be an M4 dwarf. The spectral type of the companion star determined from the SDSS spectrum, M2, is hence too early for the orbital period of the system. This suggests that the donor star is somewhat evolved, similar to the dwarf nova QZ\,Ser, a $\Porb=119.8$\,min CV which harbours a K4$\pm$2 companion star \citep{Thorstensen+02pasp}. Ultraviolet spectroscopy of SDSS\,J0924 would allow us to probe the abundance ratio of nitrogen to carbon, which is expected to be enhanced in such stars by nuclear processing \citep{Gansicke+03apj}



\section{SDSS J115207.00$+$404947.8}

\begin{figure} \includegraphics[width=0.48\textwidth,angle=0]{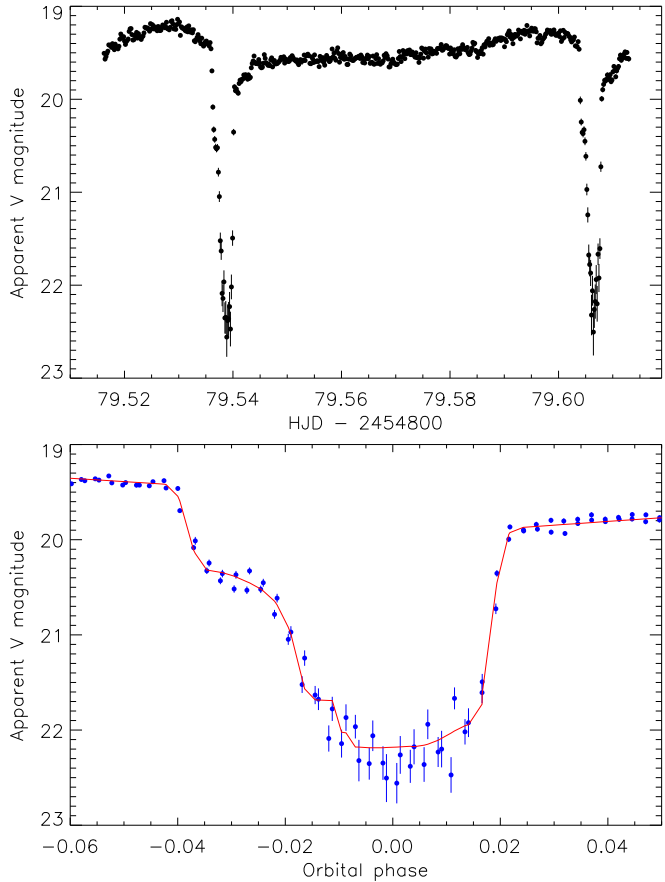} \\
\caption{\label{fig:1152} \reff{The WHT light curve} of SDSS\,J1152 against
time (upper panel) and orbital phase (lower panel). The best fit is plotted
with a solid line in the lower panel.} \end{figure}

The SDSS spectrum of SDSS\,J1152 shows a blue continuum with broad double-peaked Balmer emission features, leading  \citet{Szkody+07aj} to identify it as one of the population of SDSS CVs. The large separation of the double peaks is a telltale signature of a high orbital inclination, whilst the shallow Balmer absorption outside the emission lines is a clear sign of the underlying white dwarf.

The spectral features of SDSS\,J1152 are a reliable indicator of a high-inclination short-period CV with a low mass transfer rate (e.g.\ SDSS\,J1035; \citealt{Me+06mn} and \citealt{Littlefair+06sci}), and our observations indeed reveal two deep eclipses recurring on a period close to the minimum found in hydrogen-rich CVs (Fig.\,\ref{fig:1152}). Due to poor weather we were unable to obtain additional data, so the ephemeris is not precise:
\[ {\rm Min\,I\ (HJD}) = 2454879.5388 (2) + 0.06770 (28) \times E \]
corresponding to an orbital period of $97.5 \pm 0.4$\,min.

The phased light curve in Fig.\,\ref{fig:1152} clearly shows that the bright spot and the white dwarf are eclipsed. This makes SDSS\,J1152 a good candidate for measurement of its physical properties: we have modelled our data using {\sc lcurve} (Sect.\,\ref{sec:lcurve}) and find that the mass ratio is well-defined even though our observations are rather noisy. High-quality follow-up observations of SDSS\,J1152 are strongly encouraged.


\section{SDSS J152419.33$+$220920.1}

\begin{figure} \includegraphics[width=0.48\textwidth,angle=0]{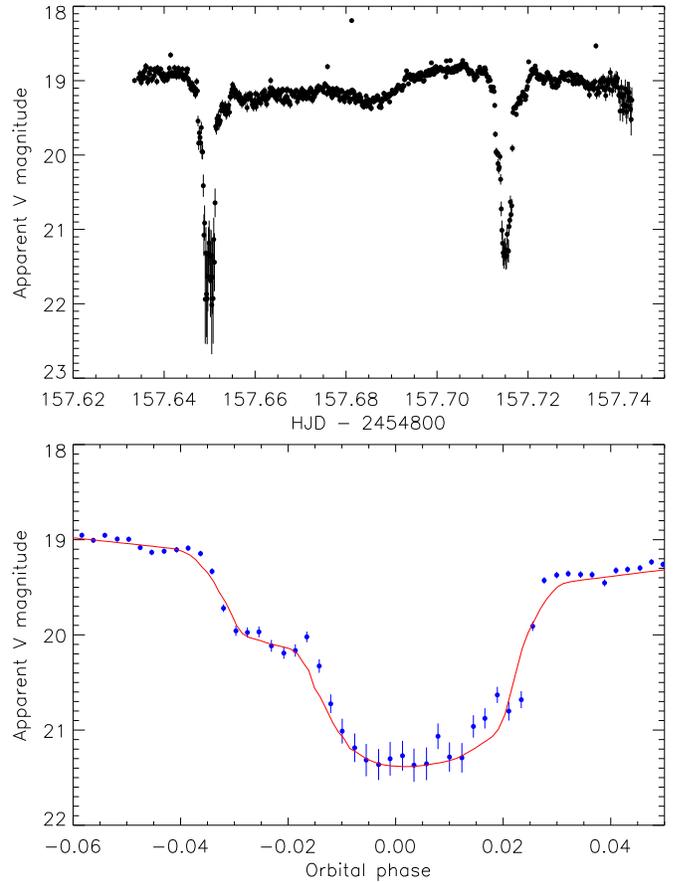} \\
\caption{\label{fig:1524} \reff{Our WHT light curve} of SDSS\,J1524 against time
(upper panel) and orbital phase (lower panel). Only the second eclipse is plotted
in the lower panel, as these data are much better than those from the first eclipse.
The best fit (solid line in the lower panel) is compromised by the slightly different
shape of the first eclipse (not shown) compared to the second eclipse.} \end{figure}

\begin{figure} \includegraphics[width=0.48\textwidth,angle=0]{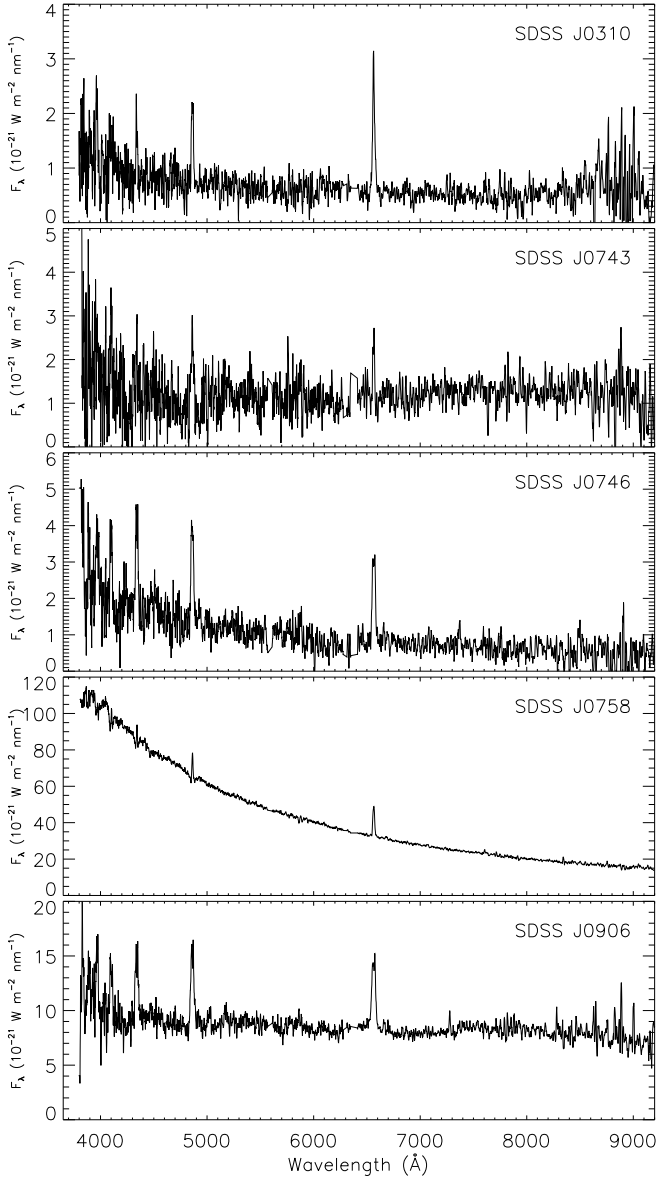} \\
\caption{\label{fig:sdssspec-n} SDSS spectra of the CVs for which we did not find periodic
brightness variations. See Fig.\,\ref{fig:sdssspec-e} caption for details.} \end{figure}

SDSS\,J1524 was discovered to be a CV by \citet{Szkody+09aj}, from an SDSS spectrum which is very similar to that of SDSS\,J1152. Szkody et al.\ stated that the deep central absorption in the centres of the double-peaked Balmer emission made it a candidate eclipsing system. Our follow-up photometry, obtained over 2.75\,hr on the night of 2009/05/06 using the WHT in service mode, confirms this suggestion. We observed two eclipses, which have a depth of 2.2\,mag and a duration of 5.6\,min and show clear structure due to the eclipse of the bright spot and of the white dwarf (Fig.\,\ref{fig:1524}). We find an ephemeris of
\[ {\rm Min\,I\ (HJD}) = 2454957.6499 (1) + 0.06500 (32) \times E \]
corresponding to an orbital period of $93.6 \pm 0.5$\,min. The measured times of minimum light are given in Table\,\ref{tab:eclipses}.

SDSS\,J1524 is strongly reminiscent of SDSS\,J1152, and our {\sc lcurve} models of the two systems are very similar (Table\,\ref{tab:lcurve}). Our modelling of the eclipses of SDSS\,J1524 was hindered by flickering, which causes the two eclipses to have different shapes. SDSS\,J1524 is, however, an excellent candidate for detailed characterisation using high-speed photometry, as flickering can be averaged out by observing several eclipses.

Since our observations were taken we have become aware that Dr.\ J.\ Patterson has independently discovered eclipses in SDSS\,J1524\footnote{See {\tt http://cbastro.org/}}. An outburst has also been detected by the {\it Catalina Sky Survey} \citep[CSS][]{Drake+09apj} using a 0.7\,m Schmidt telescope, which allows us to classify this system as a dwarf nova.


\section{Five SDSS cataclysmic variables showing no periodic brightness variations}             \label{sec:no}

\begin{figure} \includegraphics[width=0.48\textwidth,angle=0]{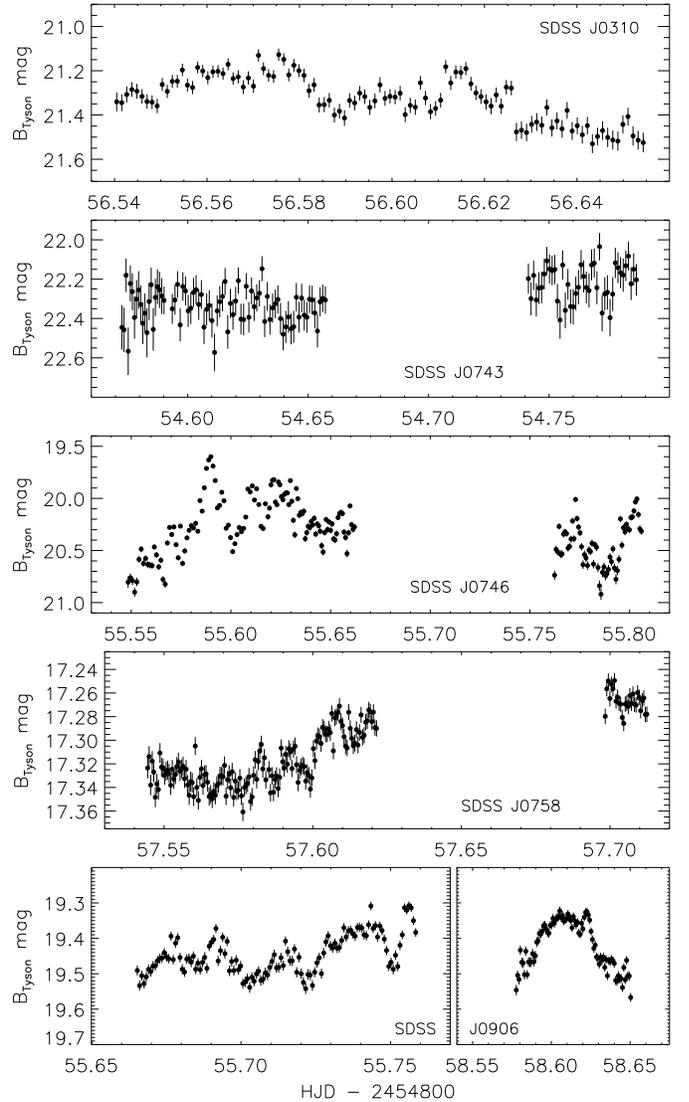} \\
\caption{\label{fig:noec} \reff{NTT light curves} of SDSS\,J0310, SDSS\,J0743,
SDSS\,J0746, SDSS\,J0758 and SDSS\,J0906.} \end{figure}

\begin{figure} \includegraphics[width=0.48\textwidth,angle=0]{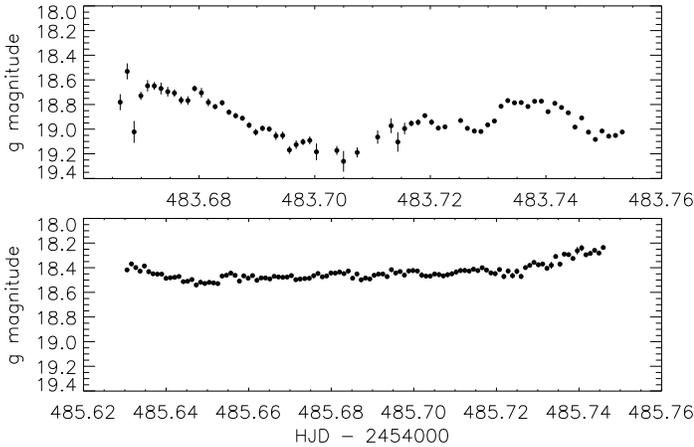} \\
\caption{\label{fig:0906}\reff{The INT light curves} of SDSS\,J0906.}\end{figure}

During our NTT observing run we obtained light curves of five CVs which did not display any clear periodic variability. Their spectra are shown in Fig.\,\ref{fig:sdssspec-n}, the log of observations is given in Table\,\ref{tab:obslog}, and our light curves are plotted in Fig.\,\ref{fig:noec}. Orbital period measurements of these objects will require spectroscopic observations, mostly with a large telescope.

We obtained a single 2.75\,hr light curve of SDSS\,J0310 \citep{Szkody+03aj}: there is a hint of a brightness variation at 76\,min but it is not significant enough to claim a detection. SDSS\,J0310 appears to outburst quite frequently. It was at magnitude $g = 15.5$ in the SDSS imaging observations, but at $g_{\rm spec} = 22.0$ in the subsequent SDSS spectroscopic observations. Szkody et al.\ then found it to be in outburst during their follow-up spectroscopic observations. An outburst was also observed by Berto Monard\footnote{{\tt vsnet-alert} number 8239}, where superhumps were visible with a period of either 0.0643(3)\,d or its one-day alias of 0.0687(3)\,d \footnote{{\tt vsnet-gcvs} number 614}.

We observed SDSS\,J0743 \citep{Szkody+06aj} on the night of 2009/01/23, for 2\,hr then another 1\,hr after a gap of 2\,hr. The light curve shows nothing out of the ordinary. One outburst has been seen for this CV in 2006 October, by the CSS \citep{Drake+09apj}, where it increased in brightness by at least 5\,mag.

SDSS\,J0746 was found to be a CV by \citet{Szkody+06aj} and is another object which was much brighter when the SDSS observed it photometrically ($g = 18.2$) than spectroscopically ($g_{\rm spec} = 21.2$). The CSS has detected two outbursts in which the object became as bright as magnitude 16 \citep{Drake+09apj}. Nakajima\footnote{{\tt vsnet-alert} number 11069}$^{,}$\footnote{{\tt vsnet-outburst} number 10022} and Maehara\footnote{{\tt vsnet-alert} number 11075} have reported a measurement of the superhump period of 0.06666(3)\,d. Our own light curve covers a total 3.75\,hr over two observing sequences on the night of 2009/01/23 (Fig.\,\ref{fig:noec}). We confirm that SDSS\,J0746 is a very variable object, but in this case we find only stochastic variations with a total amplitude of 1.3\,mag and attributable to the phenomenon of flickering \citep{Bruch92aa,Bruch00aa}.

We observed SDSS\,J0758 \citep{Szkody+09aj} for only 2.3\,hr in two segments on the night of 2009/01/25. The light curve shows some variation but our data are insufficient to be useful for period measurement. We have a short sequence of spectroscopic observations from 2008 February which indicate a period of about 80\,min and will be presented in a later paper.

SDSS\,J0906 was found to be a CV by \citet{Szkody+05aj} from an SDSS spectrum which shows moderately weak Balmer line emission. Twelve outbursts, of amplitude roughly 3\,mag, have been seen by the CSS \citep{Drake+09apj}, which makes it a frequently-outbursting system. Our 5.75\,hr of NTT observations, in three blocks over two nights, show 0.2\,mag brightness variations but nothing which is clearly periodic. We have also obtained 4.8\,hr of INT photometry of this system (Fig.\,\ref{fig:0906}) which shows similar characteristics.


\section{Summary and discussion}

\begin{table} \begin{center}
\caption{\label{tab:result} Summary of the orbital periods and CV
classification obtained for the objects studied in this work.}
\begin{tabular}{l c l} \hline
Object       & Period (min)          & Notes                      \\
\hline
SDSS\,J0310  &                       & Dwarf nova, unknown period \\
SDSS\,J0743  &                       & Dwarf nova, unknown period \\
SDSS\,J0746  &                       & Dwarf nova, unknown period \\
SDSS\,J0750  & $134.1564 \pm 0.0008$ & Eclipsing CV               \\
SDSS\,J0758  &                       & CV, no observed periodicity\\
SDSS\,J0906  &                       & Dwarf nova, unknown period \\
SDSS\,J0924  & $131.2432 \pm 0.0014$ & Eclipsing magnetic CV      \\
SDSS\,J1152  & $97.5 \pm 0.4$        & Eclipsing CV               \\
SDSS\,J1524  & $93.6 \pm 0.5$        & Eclipsing dwarf nova       \\
\hline \end{tabular} \end{center} \end{table}

We have presented NTT/EFOSC2 and WHT/ACAM time-series photometry of nine CVs identified by the SDSS, from which we find four of them to be eclipsing systems. We also present light curves of five SDSS CVs which do not show identifiable periodic phenomena. Our results are summarised in Table\,\ref{tab:result}.

SDSS\,J0750 has an orbital period of 134.1564\,min, which puts it in an interesting situation close to the lower edge of the 2--3\,hr period gap in the observed population of CVs \citep{WhyteEggleton80mn,Knigge06mn}. This period gap is commonly explained through the phenomenon of disrupted magnetic braking \citep{SpruitRitter83aa}, which requires CVs to migrate through the gap without undergoing mass transfer. A prediction of this scenario is that CVs with orbital periods of 2\,hr have secondary star masses \reff{essentially the same as} those with periods of 3\,hr; SDSS\,J0750 fits the bill perfectly as a 2\,hr CV whose donor mass can be measured \reff{precisely}.

SDSS\,J0924 has an orbital period of 131.2432\,min. It displays strong \ion{He}{II} emission and has an extremely faint or missing accretion disc, so is very likely to be a magnetic CV. As magnetic CVs evolve differently from their non-magnetic cousins \citep{WebbinkWick02mn}, SDSS\,J0924 is precluded from investigations of the disrupted magnetic braking scenario. Instead, the shape of the eclipse suggests that it will be possible to measure the radius of the WD directly. Our own observations have a cadence of 15\,s and do not resolve the ingress and egress of the WD. Light curves with a substantially higher time resolution, such as can be provided by ULTRACAM \citep{Dhillon+07mn}, will be necessary.

SDSS\,J1152 and SDSS\,J1524 are near-identical twins, for which we find orbital periods of 97.5 and 93.6 min and mass ratios of 0.14 and 0.17, respectively. Ingress and egress features of both the WD and the bright spot are clearly visible in our light curves, which makes these objects very well suited to photometric determinations of their physical properties \citep[e.g.][]{Littlefair+07mn,Littlefair+08mn}.

Five of our nine targets did not exhibit identifiable periodic brightness variations; of these SDSS\,J0746 shows a remarkable flickering activity with an amplitude of at least 1.3\,mag. Time-resolved spectroscopy is needed to measure the orbital periods of these objects, which will require 8m-class telescopes.

All four of the orbital periods measured in this work are shorter than the 2--3\,hr period gap observed in the general CV population. This is an increasingly common feature of the SDSS sample of CVs, and further evidence that previous samples are much less complete for the shorter-period faint systems than for the longer-period and generally much brighter objects \citep{Gansicke+09mn}. The SDSS CV sample is a unique window into the physical properties and evolutionary history of this faint but dominant population of CVs.


\begin{acknowledgements}

The reduced observational data presented in this work will be made available at the CDS ({\tt http://cdsweb.u-strasbg.fr/}) and at {\tt http://www.astro.keele.ac.uk/$\sim$jkt/}. JS, CMC and BTG acknowledge financial support from STFC in the form of grant number ST/F002599/1. \ We thank the anonymous referee for a timely response. Based on observations made with ESO Telescopes at the La Silla Observatory under programme ID 082.D-0605, and on observations made with the William Herschel Telescope and Isaac Newton Telescope, both operated by the Isaac Newton Group on the island of La Palma in the Spanish Observatorio del Roque de los Muchachos of the Instituto de Astrof\'{\i}sica de Canarias. The following internet-based resources were used in research for this paper: the ESO Digitized Sky Survey; the NASA Astrophysics Data System; the SIMBAD database operated at CDS, Strasbourg, France; and the ar$\chi$iv scientific paper preprint service operated by Cornell University.

\end{acknowledgements}


\bibliographystyle{aa}

\label{lastpage}

\end{document}